\pdfoutput=1

\documentclass[aps,prx,reprint,superscriptaddress,floatfix, longbibliography]{revtex4-2}

\usepackage[english]{babel}
\usepackage{graphicx}
\usepackage{amsmath}
\usepackage{amssymb}
\usepackage{bm}
\usepackage{xcolor}
\usepackage{booktabs}

\begin{document}

\title{Renormalization-group analysis of two-dimensional long-range XY models \\ with $p$-fold anisotropy: the $\sigma=2$ boundary}

\author{Xianzhi Pan}
\affiliation{Hefei National Research Center for Physical Sciences at the Microscale and School of Physical Sciences, University of Science and Technology of China, Hefei 230026, China}
\author{Chengxiang Ding}
\affiliation{School of Science and Engineering of Mathematics and Physics, Anhui University of Technology, Maanshan 243002, China }
\author{Youjin Deng}
\email{yjdeng@ustc.edu.cn}
\affiliation{Hefei National Research Center for Physical Sciences at the Microscale and School of Physical Sciences, University of Science and Technology of China, Hefei 230026, China}
\affiliation{Hefei National Laboratory, University of Science and Technology of China, Hefei 230088, China}
\affiliation{College of Physics, Guizhou University, Guiyang 550025, China}

\date{\today}

\begin{abstract}
The two-dimensional classical XY model with \(1/r^{2+\sigma}\) interactions has recently been shown to undergo, at the marginal case \(\sigma=2\), a continuous transition into a logarithmic quasi-long-range-ordered (log-QLRO) phase where the correlation decays as \(C(r) \sim (\ln r)^{-\eta_\ell}\).
We study how a \(p\)-fold field \(h_p\cos(p\theta)\) modifies this picture.
Using the physical vortex and \(p\)-fold field fugacities, we construct a unified set of renormalization-group (RG) equations that interpolate between the short-range (SR) and long-range (LR) cases through an LR coupling parameter \(\kappa\).
For \(p<4\), the small-\(h_p\) perturbative treatment breaks down and the critical behavior lies beyond the present RG treatment.
For \(p>4\), the field is irrelevant at the pure fixed point, but at low temperature the LR enhancement drives the stiffness upward: log-QLRO appears only as a finite-scale crossover, and the only transition is the continuous one from the disordered into the \(p\)-fold-ordered phase.
For \(p=4\), the fixed line shifts from the SR Ashkin--Teller form, and the transition into the fourfold-ordered phase has a nonuniversal correlation-length exponent depending on both \(\kappa\) and \(h_4\).
These results support the \(\sigma_*=2\) boundary, in contrast to Sak's criterion.
\end{abstract}

\maketitle
\section{Introduction}
\label{sec:intro}

The two-dimensional (2D) classical XY model occupies a central place in the theory of topological phase transitions. 
The Mermin--Wagner theorem~\cite{Mermin1966,Mermin1967} forbids spontaneous breaking of the continuous symmetry at any finite temperature,
yet the model exhibits the Berezinskii--Kosterlitz--Thouless (BKT) topological transition at \(T_c\), driven by the unbinding of vortex--antivortex pairs~\cite{Berezinskii1971,KT1973,Kosterlitz1974}. Below $T_c$, the low-temperature (low-\(T\)) phase displays algebraic quasi-long-range order (QLRO),
\begin{equation}
C(r)\sim r^{-\eta}, \qquad \eta=1/(2\pi K),
\label{eq:BKTcorrleation}
\end{equation}
with $K=J/k_BT$ the spin stiffness. 
Above $T_c$, free vortices proliferate and the system becomes disordered.
The renormalized stiffness has the universal limiting value $K_*=2/\pi$ at \(T_c^-\) and vanishes above $T_c$.

The interplay between the BKT mechanism and discrete symmetry-breaking fields has a long history of study. 
We consider the square-lattice XY model with a \(p\)-fold field,
\begin{equation}
\mathcal H = -J\sum_{\langle i,j\rangle}\cos(\theta_i-\theta_j)
-h_p\sum_i\cos(p\theta_i),
\label{eq:latticeH}
\end{equation}
where \(\theta_{i} \in[0,2\pi)\) is the spin phase at site \(i\), \(J>0\) is the ferromagnetic coupling, and \(K=J/k_BT\). José, Kadanoff, Kirkpatrick, and Nelson (JKKN) showed~\cite{JKKN1977} that, in the spin-wave approximation, the \(p\)-fold field has scaling dimension
 \( \lambda_p=2-p^2/(4\pi K) \) 
on the BKT fixed line. The field is therefore relevant when \(2\pi K>p^2/4\) and irrelevant when \(2\pi K<p^2/4\), while the condition for vortices to remain bound is \(2\pi K>4\). These two inequalities distinguish three regimes of critical behavior. 
For \(p<4\), the inequality \(2\pi K>p^2/4\) holds at the BKT point \(K_*=2/\pi\) and \(h_p\) is relevant; the pure planar critical behavior is destroyed and the small-\(h_p\) perturbative treatment breaks down. For \(p>4\), the two conditions can be satisfied simultaneously, and the stability window
\begin{equation}
4<2\pi K<p^2/4
\end{equation}
has finite width. Inside this window both the vortex fugacity and the \(p\)-fold field are irrelevant, so that the isotropic planar critical behavior is observable. For \(p=4\), the two conditions coincide at \(2\pi K=4\): the window collapses to zero width, \(h_4\) is marginal at the BKT point, and JKKN argued that this generates an Ashkin--Teller/Baxter fixed line \(y=\pm y_4\) with continuously varying exponents~\cite{AshkinTeller1943,Baxter1971,Baxter1982}.

The JKKN picture has been confirmed numerically in both the clock limit and at finite anisotropy. Clock-model simulations for \(q\ge5\) exhibit the two BKT transitions with \(\eta_1=1/4\) and \(\eta_2=4/q^2\)~\cite{ChenHouFangDeng2022}. Finite-anisotropy simulations with fourfold, fivefold, and sixfold fields~\cite{Rastelli2004Fields,Itakura2001,Butt2023} are consistent with the JKKN stability window and the two-transition scenario, although the convergence near the lower transition is slow. For \(p=4\), early weak-field simulations reported an intermediate BKT phase~\cite{Rastelli2004Weak}, but subsequent work showed that this is a finite-size crossover~\cite{Taroni2008}; large-scale simulations support a single continuous transition with a nonuniversal exponent varying with \(h_4\)~\cite{Yen2026}.

Recently, the fate of the BKT physics under algebraic LR interactions \(1/r^{2+\sigma}\) has attracted considerable attention, motivated by quantum simulators with tunable interaction ranges~\cite{Monroe2021,Britton2012,Chen2023,Bonifacio2024,Yan2013}. The phase diagram of the 2D LR XY model has been debated, with three main scenarios~\cite{Giachetti2021,Walther2026,Xiao2025LRXY,Yao2025LRXY}.
A combination of large-scale simulations~\cite{Xiao2025LRXY,Yao2025LRXY}, the Gaussian spin-wave analysis and the extended RG treatment~\cite{Pan2026logXY} now supports a crossover at \(\sigma_*=2\): for \(\sigma<2\) the LR interaction is relevant and drives true LRO, while for \(\sigma>2\) standard BKT physics is recovered.
At \(\sigma=2\), the low-\(T\) phase is a logarithmic QLRO (log-QLRO), in which the spin correlation decays as 
\begin{equation}
C(r)\sim(\ln r)^{-\eta_\ell},
\qquad
\eta_\ell=1/(2\pi\kappa),
\label{eq:C_log-QLRO}
\end{equation}
where \(\kappa \propto1/T\) is the reduced LR coupling.
The log-QLRO phase is distinct from both the true LRO for \(\sigma<2\) and the algebraic QLRO for \(\sigma>2\).
The logarithmic form originates from the infrared stiffness kernel \( \gamma(k)\simeq\kappa k^2\ln(1/k)\),
which destroys true LRO and simultaneously suppresses vortex proliferation: the vortex-vortex energy grows as \((\ln r)^2\), so free vortices are super-logarithmically confined.
Instead of a BKT transition, the transition out of the log-QLRO phase is a genuine continuous transition with a power-law diverging correlation length.
Yet it does not fit the Ginzburg--Landau--Wilson (GLW) paradigm, since the low-\(T\) phase lacks true LRO and the correlation-length exponent \(\nu\) is nonuniversal, depending explicitly on \(\kappa\).
These predictions have been directly tested in large-scale Monte Carlo simulations~\cite{Pan2026logXY}, 
including the spin-wave identity \(2\pi\kappa_R\eta_\ell=1\), where $\kappa_R $ is the renormalized LR coupling strength.
The log-QLRO phase has also been observed in the \(\sigma=2\) Heisenberg model~\cite{Xiao2025Heisenberg}.

In this paper we study how the JKKN symmetry-breaking-field analysis is modified at the marginal case \(\sigma=2\).
Using the physical vortex and \(p\)-fold field fugacities, we construct a unified set of lowest-order RG equations that interpolate between 
the SR \(\sigma>2 \) (\(\kappa=0\)) and LR \(\sigma=2\) (\(\kappa>0\)) cases. 
As in the SR case, the \(p\)-fold field is relevant, marginal, or irrelevant at the pure fixed point (\(h_p=0, \kappa>0\)), according to whether \(p<4\), \(p=4\), or \(p>4\).
For \(p<4\), the \(p\)-fold field is relevant at the pure fixed point, so the critical behavior lies beyond the present RG framework.

\begin{figure*}[t]
\centering
\includegraphics[width=0.97\textwidth]{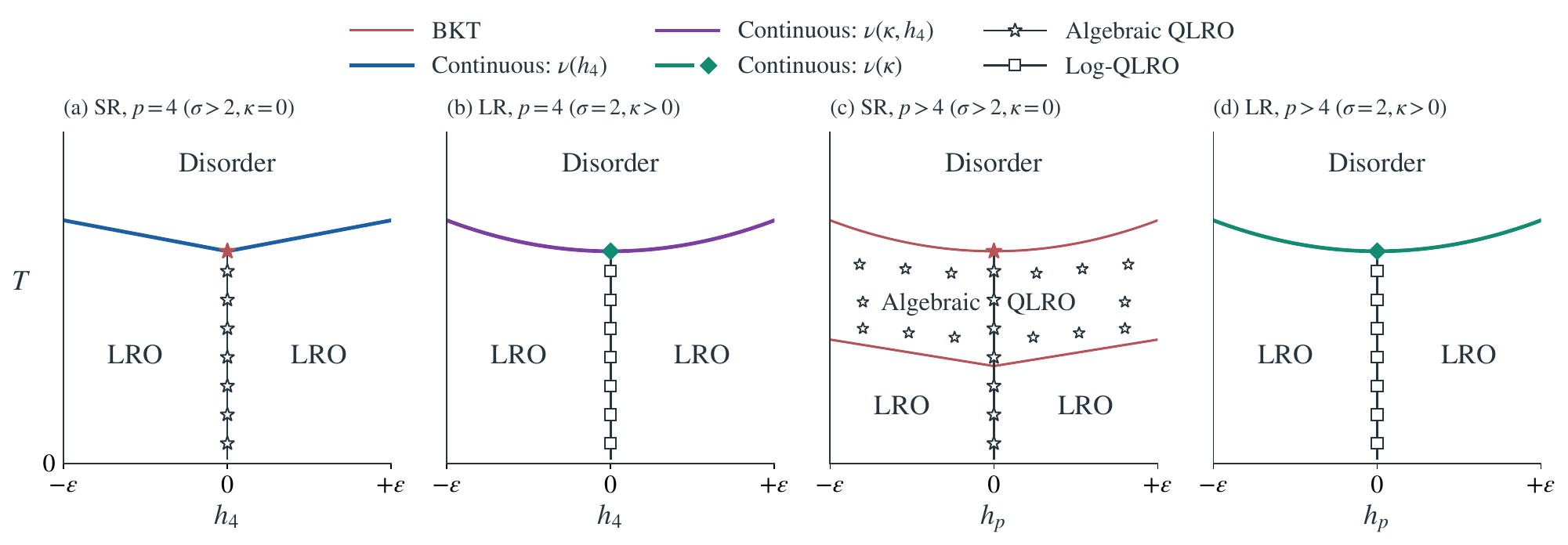}
\caption{Schematic phase diagrams in the $(h_p, T)$ plane. 
(a) SR, $p=4$ (\(\sigma>2, \kappa=0\)): along $h_4=0$, an algebraic-QLRO line terminates at the BKT endpoint, whereas for $h_4\neq 0$ a single continuous transition with $\nu(h_4)$ separates disorder from fourfold LRO. 
(b) LR, $p=4$ (\(\sigma=2, \kappa>0\)): along $h_4=0$, a log-QLRO line terminates at the pure-LR continuous point with $\nu(\kappa)$, while for $h_4\neq 0$ the continuous transition has $\nu(\kappa,h_4)\to\nu(\kappa)$ as $h_4\to 0$. 
(c) SR, $p>4$ (\(\sigma>2, \kappa=0\)): for $h_p\neq 0$, an intermediate algebraic-QLRO phase sits between two BKT transitions, whereas strict $h_p=0$ has only the upper BKT transition. 
(d) LR, $p>4$ (\(\sigma=2, \kappa>0\)): the intermediate phase reduces to a finite-scale log-QLRO crossover, leaving a single continuous transition controlled explicitly by $\nu(\kappa)$. Boundary shapes and temperatures are schematic.}
\label{fig:phaseproposal}
\end{figure*}
Figure~\ref{fig:phaseproposal} summarizes the schematic \((h_p,T)\) phase diagrams for the SR and LR models with \(p=4\) and \(p>4\).
For $\sigma>2$, the model is effectively SR, and the phase diagrams [Figs.~\ref{fig:phaseproposal}(a) and~(c)] are qualitatively the same as the JKKN models with nearest-neighbor interactions.
For $\sigma=2$, the low-\(T\) phase of the pure XY model is no longer an algebraic-QLRO phase but a log-QLRO phase.
For \(p>4\) [Fig.~\ref{fig:phaseproposal}(d)], the \(p\)-fold field is irrelevant at the pure critical fixed point.
However, the log-QLRO regime of the pure XY model is unstable against an arbitrarily small \(h_p\): 
at low temperature the LR stiffening drives \(K\) monotonically upward, and when \(K\) reaches \(K_p=p^2/(8\pi)\) the \(p\)-fold field becomes relevant and locks the system into the \(p\)-fold LRO phase. Thus, for any finite \(p>4\), log-QLRO exists only as a finite-scale crossover, not as a thermodynamic phase.
Consequently, there is a single thermodynamic transition at \(T_c\), from the disordered into a \(p\)-fold LRO phase, 
and the critical behavior at \(T_c\) is controlled by the pure fixed point with \(1/\nu \simeq \sqrt{2\pi\kappa}\).
As \(p\to\infty\), \(K_p\to\infty\), the locking scale diverges, and the pure XY log-QLRO phase is recovered.

The marginal case \(p=4\) is special because the \(p\)-fold edge coincides with the pure critical stiffness $K_*=2/\pi$.
In the SR \(\sigma >2\) (\(\kappa=0\)) case [Fig.~\ref{fig:phaseproposal}(a)], \(h_4\) is marginal at the BKT point 
(marked by a filled star) and generates the Ashkin--Teller/Baxter fixed line \(K=2/\pi\), \(y=\pm y_4\), along which \(\nu \sim 1/|h_4|\) for small $h_4$~\cite{JKKN1977}.
In the \(\sigma=2\) case [Fig.~\ref{fig:phaseproposal}(b)], the logarithmic stiffness kernel shifts this line to \( K=2/\pi \), \( y^2-y_4^2 = \kappa /(16\pi) \).
This line emanates from the pure LR fixed point \((K_*,y_*,0)\) with \(y_*=\sqrt{\kappa/(16\pi)}\), rather than from the BKT point \((K_*,0,0)\) as in the SR case.
The exponent $\nu$ is nonuniversal and depends explicitly on \(\kappa\) and \(h_4\);
it does not diverge but takes a finite value $\simeq 1/\sqrt{2\pi\kappa}$ at the $h_4=0$ endpoint (marked by a filled diamond).
The low-\(T\) phase for \(h_4\neq0\) is fourfold LRO.

These RG analyses show that \(\sigma=2\) is the SR--LR boundary, irrespective of whether the $p$-fold field is present or not.
The fixed-point structure changes: as \(\sigma\) is reduced from above 2 to the marginal value 2, the fixed point is displaced.
If Sak's criterion were correct~\cite{Sak1973}, this displacement would occur for \(\sigma<2\).

The remainder of this paper is organized as follows. After reviewing the three known limits of the model in Sec.~\ref{sec:review}, we combine them in Sec.~\ref{sec:phases} into a unified set of lowest-order RG equations for the \(\sigma=2\) model with a \(p\)-fold field, and analyze the resulting phase diagram for the \(p<4\), \(p=4\), and \(p>4\) regimes. In Sec.~\ref{sec:conclusions} we summarize our results, discuss their implications, and comment on the limitations of the adiabatic approximation.

\section{Short-range, long-range, and $p$-fold limits}
\label{sec:review}
In this section we review the three known limits on which our analysis is built: the pure SR XY model, the SR XY model with a \(p\)-fold field, and the pure LR \(\sigma=2\) XY model. 
The first two are well established, while the third is more recent and supplies the LR stiffening \(\kappa\) that, together with the \(p\)-fold fugacity, enters the unified equations of Sec.~\ref{sec:phases}.
We present these limits in some detail so that the paper is self-contained and the origin of each term in the unified equations is transparent; a reader already familiar with these results may skip directly to Sec.~\ref{sec:phases}.

\subsection{Pure short-range XY model}
\label{sec:reviewSR}

We start from the SR square-lattice Hamiltonian described by Eq.~\eqref{eq:latticeH}, but without the $p$-fold symmetry-breaking field ($h_p=0$).
At low temperature, the phase differences across neighboring sites are small, and the continuum Gaussian Hamiltonian reads
\[
\mathcal H_{\rm SR}\simeq\frac{J}{2}\int d^2r\,(\nabla\theta)^2.
\]
The Gaussian spin-wave sector is obtained by keeping only this quadratic term. 
The spin correlation decays algebraically as in Eq.~\eqref{eq:BKTcorrleation},
so the low-\(T\) phase is algebraically quasi-long-range ordered (QLRO).

The spin-wave sector is not the whole story: vortices are topological defects that cost a logarithmically growing energy.
A single vortex in a system of linear size \(L\) costs \(\pi K\ln(L)\), while its configurational entropy is \(2\ln(L)\). 
The free energy \(F_v=(\pi K-2)\ln(L)\) is positive for \(2\pi K>4\), so free vortices are absent; at \(K_*=2/\pi\) the free energy vanishes and vortices unbind. This is the BKT topological transition.

To describe the competition between spin waves and vortices, we introduce the physical vortex fugacity \(y\) as
\begin{equation}
y=(1/\sqrt2) y_0^{\rm bare}e^{-\pi^2K/2},
\label{eq:yphysSR}
\end{equation}
where the exponential factor converts the bare dual Coulomb-gas parameter, \(y_0^{\rm bare}\), into a physical fugacity.
The factor \(1/\sqrt2\) is a normalization convention. In terms of \(y\), the RG equations read
\begin{equation}
\frac{dK}{d\ell}=-4\pi^3K^2y^2,
\qquad
\frac{dy}{d\ell}=(2-\pi K)y,
\label{eq:BKT}
\end{equation}
with \(\ell=\ln b\). The first equation says that vortex pairs soften the stiffness; the second says that vortices are relevant for \(2\pi K<4\) and irrelevant for \(2\pi K>4\). 
The fixed line is
\begin{equation}
y=0,
\qquad
2\pi K>4,
\label{eq:BKTfixed}
\end{equation}
which corresponds to the vortex vacuum: free vortices are absent ($y=0$), and only the Gaussian spin-wave sector remains. Along this line the correlation exponent varies continuously, \(\eta=1/(2\pi K)\). The endpoint \(2\pi K=4\) of this line is the BKT critical point, where \(\eta=1/4\) and the correlation length diverges exponentially, \(\xi\sim\exp(b/\sqrt{t})\), as the transition is approached from the high-\(T\) side (\(t=(T-T_c)/T_c>0\)).

\subsection{Short-range XY model with a \(p\)-fold field}
\label{sec:reviewSRp}

The SR XY model with a \(p\)-fold field is given by Eq.~\eqref{eq:latticeH}.
In the spin-wave approximation, the field is a perturbation with scaling dimension~\cite{JKKN1977}
\begin{equation}
\lambda_p=2-p^2/(4\pi K)
\label{eq:lambdap}
\end{equation}
on the BKT fixed line~\eqref{eq:BKTfixed},
originating from the power-law correlation of the operator \(e^{ip\theta}\),
\( \langle e^{ip\theta(r)}e^{-ip\theta(0)}\rangle \sim r^{-p^2/(2\pi K)} \).
The field is relevant for \(2\pi K>p^2/4\), marginal at \(2\pi K=p^2/4\), and irrelevant for \(2\pi K<p^2/4\).

To go beyond the scaling dimension, we need the RG equations for the coupled vortex and \(p\)-fold fugacities. 
The \(p\)-fold fugacity is defined analogously to Eq.~\eqref{eq:yphysSR},
\begin{equation}
y_p=(1/\sqrt2) y_p^{\rm bare}e^{-p^2K^{-1}/8},
\label{eq:ypphys}
\end{equation}
and the RG equations, valid for general \(p\), read
\begin{equation}
\frac{dK}{d\ell}
=
-4\pi^3K^2y^2
+\pi p^2y_p^2,
\label{eq:JKKNK}
\end{equation}
\begin{equation}
\frac{dy}{d\ell}=(2-\pi K)y,
\qquad
\frac{dy_p}{d\ell}
=
\left(2-\frac{p^2}{4\pi K}\right)y_p.
\label{eq:JKKNy}
\end{equation}
The first term in Eq.~\eqref{eq:JKKNK} is the vortex softening; the second term is the stiffening due to the \(p\)-fold field, which prefers a definite orientation and therefore suppresses phase fluctuations. The flow of \(y\) is controlled by \(2-\pi K\), as in the pure SR case, while the flow of \(y_p\) is controlled by the same scaling dimension as in Eq.~\eqref{eq:lambdap}.

Combining the condition \(2\pi K>4\) for bound vortices with the condition \(2\pi K<p^2/4\) for an irrelevant \(p\)-fold field, we obtain the stability window
\begin{equation}
4<2\pi K< p^2/4 .
\label{eq:window}
\end{equation}
Inside the window, both vortices and the \(p\)-fold field are irrelevant, and the pure planar critical behavior is observable. The window has finite width only for \(p>4\); for \(p=4\) it collapses; for \(p<4\) it is empty. 

{\em The \(p=4\) case.---} The field is marginal at the BKT point \(K_*=2/\pi\). Setting the right-hand sides of Eqs.~\eqref{eq:JKKNK}--\eqref{eq:JKKNy} to zero gives
\begin{equation}
K=2/\pi,
\qquad
y^2=y_4^2,
\label{eq:ATline}
\end{equation}
i.e., \(y=\pm y_4\). This is a line of fixed points rather than a single critical point~\cite{JKKN1977}.
The endpoint \(y=y_4=0\) of this line is the BKT critical point; away from the endpoint, the finite fugacities shift the exponents, producing the continuously varying behavior characteristic of the Ashkin--Teller line.

To obtain the correlation-length exponent \(\nu\), we linearize Eqs.~\eqref{eq:JKKNK}--\eqref{eq:JKKNy} around the line \(y=y_4=u\), with \(K=2/\pi+\delta K\) and \(v=\delta y-\delta y_4\). One finds
\begin{equation}
\frac{d}{d\ell}
\begin{pmatrix}
\delta K\\
v
\end{pmatrix}
=
\begin{pmatrix}
-16\pi^2u^2 & -32\pi u\\
-2\pi u & 0
\end{pmatrix}
\begin{pmatrix}
\delta K\\
v
\end{pmatrix}.
\end{equation}
The relevant eigenvalue is \(\lambda_+\simeq8\pi u\) for small \(u\), so
\begin{equation}
1/\nu\simeq 8\pi y_4 \sim |h_4| \qquad(h_4\to0).
\label{eq:nuSRp4}
\end{equation}
Thus \(\nu\) diverges as \(h_4\to0\), and the low-\(T\) phase for \(h_4\neq0\) is fourfold LRO, not algebraic QLRO.

{\em The \(p>4\) case.---} The window~\eqref{eq:window} has finite width. Inside the window, the \(p\)-fold field is irrelevant at the BKT fixed line and the algebraic QLRO phase is stable. 
The two boundaries of the window correspond to two BKT transitions: the high-\(T\) transition at \(K_*=2/\pi\), where vortices unbind, 
and the low-\(T\) transition at \(K_p =p^2/(8\pi)\), where the \(p\)-fold field becomes marginal.
At these two points the anomalous dimension $\eta=1/(2\pi K)$ takes the values
\[
\eta_1=1/4,
\qquad
\eta_2=4/p^2.
\]
This two-transition scenario has been confirmed numerically in the \(p\)-state clock model, which corresponds to the \(h_p\to\infty\) limit. High-precision Monte Carlo and tensor-network studies have established for \(p\ge5\) the two BKT transitions with \(\eta_1=1/4\) and \(\eta_2=4/p^2\), separated by an algebraic QLRO phase~\cite{ChenHouFangDeng2022}.

For \(p<4\), the window~\eqref{eq:window} is empty: the \(p\)-fold field is already relevant at the BKT point. 
The small-\(h_p\) perturbative treatment breaks down in the infrared, and the critical behavior lies beyond the present RG framework. 
A separate analysis is required.

\subsection{Pure long-range \(\sigma=2\) XY model}
\label{sec:reviewLR}

We now turn to the LR XY model. On a square lattice, the Hamiltonian is
\begin{equation}
\mathcal H
=
-\sum_{i<j}J_{ij}\cos(\theta_i-\theta_j),
\qquad
J_{ij}=J/r_{ij}^{2+\sigma},
\label{eq:HLRlattice}
\end{equation}
where \(r_{ij}\) is the distance between sites \(i\) and \(j\), and the sum runs over all pairs. 
In the continuum limit, Eq.~\eqref{eq:HLRlattice} becomes
\begin{equation}
\mathcal H
=
-\frac{1}{2}\int d^2x\,d^2y\,\frac{J }{|x-y|^{2+\sigma}}
\cos[\theta(x)-\theta(y)],
\label{eq:HLRcont}
\end{equation}
where the \(1/2\) avoids double counting of the pairs.

The Gaussian spin-wave action is no longer local. Transforming to momentum space, the LR contribution to the stiffness kernel is
\begin{equation}
\begin{split}
\gamma_{\rm LR}(k)
&=
K\int d^2r\,\frac{1-\cos({\bf k}\cdot{\bf r})}{r^{2+\sigma}}\\
&=
2\pi K k^\sigma\int_{k}^{\infty}\frac{du}{u^{1+\sigma}}
[1-J_0(u)],
\end{split}
\label{eq:gammaLR}
\end{equation}
where \(K=J/(k_BT)\) and \(J_0\) is the Bessel function. For small \(k\), the asymptotic behavior depends on \(\sigma\):
\begin{equation}
\gamma(k)\simeq
\begin{cases}
B_\sigma k^2, & \sigma>2,\\[1mm]
\kappa k^2\ln(1/k), & \sigma=2,\\[1mm]
A_\sigma k^\sigma, & 0<\sigma<2,
\end{cases}
\label{eq:kernel3}
\end{equation}
where \(B_\sigma\), \(A_\sigma\), and \(\kappa=\pi J /(2k_BT)\) are positive constants. 
For \(\sigma>2\), the LR contribution is subleading and only renormalizes the coefficient of 
the SR kernel \(\sim k^2\), and standard BKT physics is recovered. 
For \(\sigma<2\), the LR kernel $k^\sigma$ gives rise to true LRO. 
The case \(\sigma=2\) sits precisely between the two: the kernel is \(k^2\ln(1/k)\), 
harder than the SR \(k^2\) but softer than the LR \(k^\sigma\).

We focus on \(\sigma=2\). The low-\(T\) phase is a log-QLRO phase~\cite{Pan2026logXY}, 
in which the spin correlation decays as a power of the logarithm of distance, described by Eq.~\eqref{eq:C_log-QLRO}.
This decay is slower than any power law but faster than a constant, so the phase is distinct from both the true LRO for \(\sigma<2\) and the algebraic QLRO  for \(\sigma>2\). 
The logarithmic kernel suppresses vortex proliferation. A single vortex costs an energy
\[
E_v(L)\sim\frac{\pi\kappa}{2}(\ln L)^2,
\]
which dominates the configurational entropy \(2\ln L\). Free vortices are therefore super-logarithmically confined, and the standard BKT unbinding mechanism is absent.

To describe the transition out of the log-QLRO phase, we use the adiabatic approximation.
The key observation is that the \(\sigma=2\) kernel differs from the SR kernel only by the factor \(\ln(1/k)\). Under a Wilsonian rescaling \(k'=bk\), this factor generates an additive contribution to the stiffness, \(\delta K=\kappa\ln b\), accumulated over each momentum shell. Absorbing it into a running stiffness,
\begin{equation}
K(\ell)=K_0+\kappa\ell,
\qquad
\ell=\ln b,
\label{eq:runningK}
\end{equation}
we obtain \(dK/d\ell=\kappa\) from the LR sector. The vortex sector is unchanged, because vortices still interact logarithmically with the running stiffness \(K(\ell)\). This gives the extended BKT equations~\cite{Pan2026logXY}
\begin{equation}
\frac{dK}{d\ell}=\kappa-4\pi^3K^2y^2,
\qquad
\frac{dy}{d\ell}=(2-\pi K)y.
\label{eq:LRBKT}
\end{equation}
The nontrivial fixed point is
\begin{equation}
K_*=2/\pi,
\qquad
y_*=\sqrt{\kappa/(16\pi)}.
\label{eq:LRfixed}
\end{equation}
Unlike the SR case, where \(y=0\) is a fixed line for all \(2\pi K>4\), the LR stiffening \(\kappa>0\) makes \(y=0\) non-fixed; the only nontrivial fixed point is the isolated point~\eqref{eq:LRfixed}. Linearizing around \((K_*,y_*)\) gives the relevant eigenvalue \(\lambda_+\simeq\sqrt{2\pi\kappa}\).
Because the fixed point has a finite \(y_*\) and a nonzero relevant eigenvalue, the transition is a genuine continuous transition with a power-law divergent correlation length,
\begin{equation}
\xi\sim|t|^{-\nu},
\qquad
1/\nu\simeq \sqrt{2\pi\kappa}.
\label{eq:nuLR}
\end{equation}
The exponent is nonuniversal and depends explicitly on $\kappa$. The critical stiffness is still \(K_*=2/\pi\),
so \(\eta\) remains \(1/4\) at lowest order. The finite \(y_*\) is the structural difference from the SR case: the LR stiffening balances the vortex softening, leading to a single fixed point rather than a fixed line.

The adiabatic approximation becomes increasingly accurate as temperature is lowered.
The log-QLRO phase and the continuous transition have been directly tested in large-scale simulations: the measured squared magnetization follows \(M^2\sim(\ln L)^{-\eta_\ell}\), the correlation-length ratio grows as \((\xi/L)^2\simeq\kappa_R\ln L\), and the independently extracted \(\kappa_R\) and \(\eta_\ell\) satisfy the spin-wave identity \(2\pi\kappa_R\eta_\ell=1\).
In the critical region, vortex fluctuations and the flow of \(\kappa\) modify the quantitative values of the fixed-point parameters. 

\begin{figure}[t]
\centering
\includegraphics[width=0.98\columnwidth]{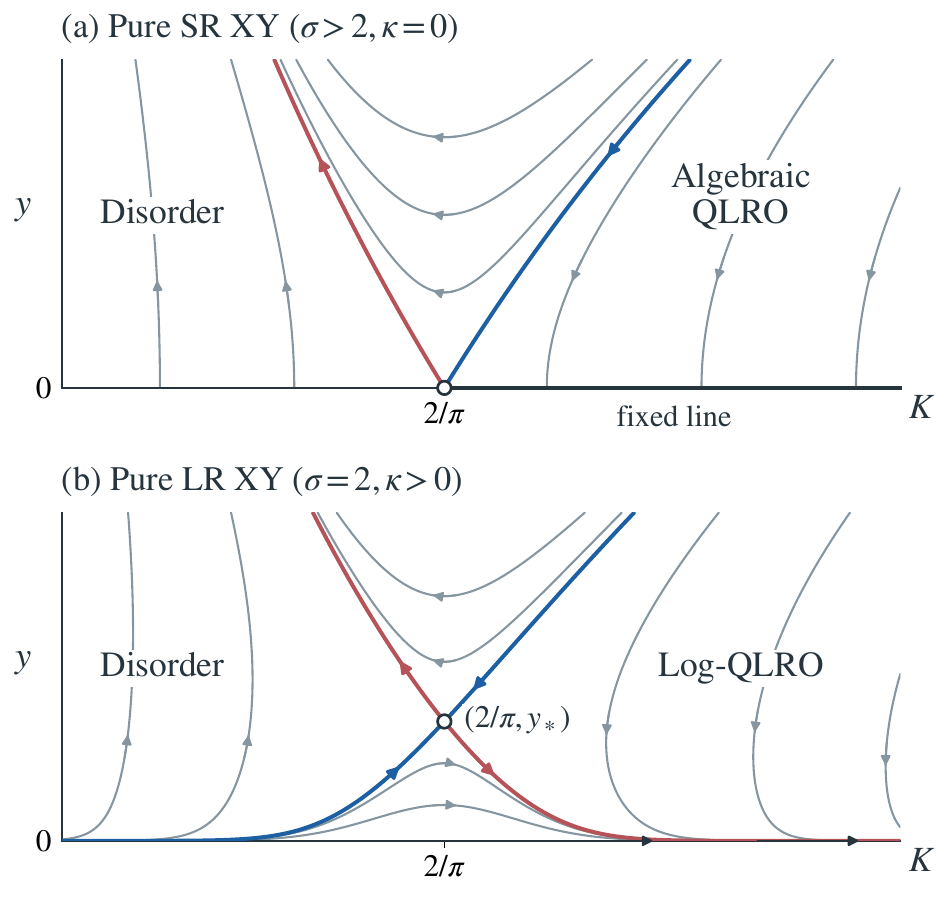}
\caption{Numerical pure XY RG flows in the $(K,y)$ plane, obtained from Eqs.~\eqref{eq:BKT} and~\eqref{eq:LRBKT}. (a) SR ($\sigma>2$): the critical separatrix approaches the BKT endpoint $(2/\pi,0)$ of the Gaussian fixed line, separating the disordered and algebraic-QLRO phases. (b) LR ($\sigma=2$) with $\pi\kappa=0.04$: the endpoint becomes a saddle fixed point $(2/\pi,y_*)$ with $y_*=\sqrt{\kappa/(16\pi)}$, and low-temperature flows characterize log-QLRO with $y\to0$ and $K$ growing logarithmically with length scale. Blue and red curves denote the critical separatrix and outgoing flows from the fixed point, respectively; gray curves are representative flows. Arrows indicate increasing RG scale.}
\label{fig:purexyflowsnew}
\end{figure}

The contrasting RG flows of the pure SR and LR XY models are shown in Fig.~\ref{fig:purexyflowsnew}. In the SR case [Fig.~\ref{fig:purexyflowsnew}(a)], the critical separatrix terminates at the BKT endpoint \((2/\pi,0)\) of the Gaussian fixed line, separating the disordered and algebraic-QLRO phases. At \(\sigma=2\) [Fig.~\ref{fig:purexyflowsnew}(b)], balancing the LR stiffening \(\kappa>0\) requires finite vortex softening at the fixed point, lifting it off the \(y=0\) axis to a finite-vortex-fugacity saddle point \((2/\pi,y_*)\). Its critical separatrix separates the disordered phase from the low-\(T\) log-QLRO regime, where \(y\to 0\) and \(K\) grows logarithmically with length scale.

\section{Unified RG analysis of $\sigma=2$ with $p$-fold anisotropy}
\label{sec:phases}

We now combine the three limits reviewed in Sec.~\ref{sec:review} into a unified RG scheme for the XY model with both a \(\sigma=2\) LR interaction and a \(p\)-fold field. On the square lattice, the Hamiltonian reads
\begin{equation}
\mathcal H
= -\sum_{i<j}J_{ij}\cos(\theta_i-\theta_j)
-h_p\sum_i\cos(p\theta_i),
\label{eq:Hfull}
\end{equation}
with $J_{ij}=J/r_{ij}^{4}$.
The first term describes the LR interaction at \(\sigma=2\) and the second term is the \(p\)-fold field.

Equation~\eqref{eq:Hfull} contains the above three limits. Setting \(h_p=0\) and \(\sigma>2\) recovers the pure SR model. Keeping \(h_p\neq0\) but setting \(\sigma>2\) recovers the SR model with a \(p\)-fold field. Setting \(h_p=0\) but keeping \(\sigma=2\) recovers the pure LR model. 
Combining the SR and LR equations~\eqref{eq:JKKNK}, \eqref{eq:JKKNy} and~\eqref{eq:LRBKT}, we obtain the RG equations for the full model 
\begin{equation}
\frac{dK}{d\ell}
=
\kappa
-4\pi^3K^2y^2
+\pi p^2y_p^2,
\label{eq:unifiedK}
\end{equation}
\begin{equation}
\frac{dy}{d\ell}=(2-\pi K)y,
\qquad
\frac{dy_p}{d\ell}
=
\left(2-\frac{p^2}{4\pi K}\right)y_p,
\label{eq:unifiedy}
\end{equation}
with the physical fugacities \(y\) and \(y_p\) defined in Eqs.~\eqref{eq:yphysSR} and~\eqref{eq:ypphys}.

\begin{figure}[t]
\centering
\includegraphics[width=0.98\columnwidth]{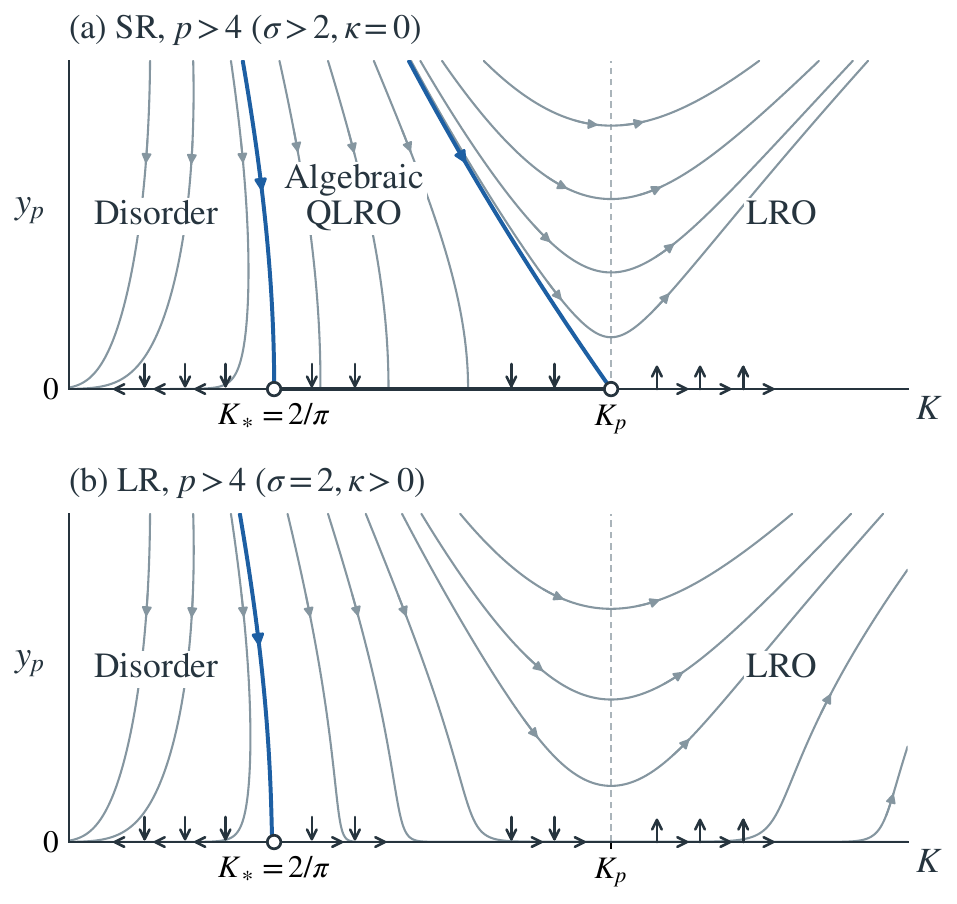}
\caption{Numerical RG flows for $p>4$ (illustrated using $p=6$), obtained from Eqs.~\eqref{eq:unifiedK}--\eqref{eq:unifiedy} in the three-dimensional $(K,y,y_p)$ space and orthogonally projected for display onto the $y=y_*$ [$(K,y_p)$] plane (with $y_*=0$ for SR and $y_*=\sqrt{\kappa/(16\pi)}$ for LR); $y$ remains dynamical during integration. (a) SR ($\sigma>2$): the BKT endpoints $K_*=2/\pi$ and $K_p=p^2/(8\pi)$ (vertical dashed line) bound the stable Gaussian fixed segment, separating the disordered, algebraic-QLRO, and LRO phases. (b) LR ($\sigma=2$) with $\pi\kappa=0.04$: the pure-LR saddle fixed point $(K_*,y_*,0)$ projects to $(K_*,0)$, and on the ordered side a transient log-QLRO regime precedes locking to LRO.}
\label{fig:p6projectednew}
\end{figure}

Each term in Eq.~\eqref{eq:unifiedK} has a clear physical meaning. The constant \(\kappa\) is the LR stiffening: the \(1/r^4\) interaction enhances the stiffness. The term \(-4\pi^3K^2y^2\) is the vortex softening: vortex pairs reduce the stiffness. The term \(+\pi p^2y_p^2\) is the \(p\)-fold stiffening: the field prefers a definite orientation and suppresses phase fluctuations. The flow of \(y\) is controlled by \(2-\pi K\), as in the SR and pure LR cases. The flow of \(y_p\) is controlled by \(2-p^2/(4\pi K)\), which is the same scaling dimension as in Eq.~\eqref{eq:lambdap}.

The LR term \(\kappa\) is marginal at \(\sigma=2\) and does not flow at lowest order, so Eqs.~\eqref{eq:unifiedK}--\eqref{eq:unifiedy} describe a family of three-dimensional flows. Before analyzing these flows, it is worth noting a duality property of the SR limit. For \(\kappa=0\), the equations are invariant under the exchange
\begin{equation}
y\leftrightarrow y_p,
\qquad
K\rightarrow p^2/(4\pi^2K),
\label{eq:duality}
\end{equation}
which is the field-theoretic counterpart of the Kramers--Wannier duality of the clock model. This duality maps the two edges of the SR stability window, \(2\pi K=4\) and \(2\pi K=p^2/4\), onto each other, and it is responsible for the symmetric role played by the vortex and \(p\)-fold fugacities in the SR Ashkin--Teller line. The LR stiffening \(\kappa>0\) breaks this duality explicitly: the constant term in Eq.~\eqref{eq:unifiedK} is not invariant under Eq.~\eqref{eq:duality}. 

The RG equations~\eqref{eq:unifiedK}--\eqref{eq:unifiedy} are the lowest-order terms of an adiabatic expansion in which $\kappa$ is treated as scale independent. In the critical region, the vortex fugacity is no longer exponentially small and \(\kappa\) itself may flow,
and the quantitative values of the critical parameters are modified.
Nevertheless, the qualitative conclusions drawn in this section are expected to survive. 

At the pure LR fixed point \((K,y,y_p)=(K_*,y_*,0)\), with \(K_*,y_*\) given by Eq.~\eqref{eq:LRfixed},
the three regimes are also classified by whether \(p<4\), \(p=4\), or \(p>4\).

\subsection{The marginal case \(p=4\)}
\label{sec:p4}

For \(p=4\), the \(p\)-fold edge coincides with the pure critical stiffness, \(K_4=p^2/(8\pi)=2/\pi=K_*\).
This coincidence is the source of the special structure of the \(p=4\) case. 
In the SR \(\sigma>2\) case, the field is marginal at the BKT point and generates the Ashkin--Teller fixed line~\eqref{eq:ATline}.
The correlation-length exponent $\nu$ in Eq.~\eqref{eq:nuSRp4} is nonuniversal, depending on $h_4$, and the low-\(T\) phase is fourfold LRO.

For \(\sigma=2\) (i.e., \(\kappa>0\)), the fixed-point conditions again require \(K=2/\pi\). Substituting into Eq.~\eqref{eq:unifiedK} gives
\begin{equation}
\kappa-16\pi y^2+16\pi y_4^2=0,
\label{eq:LRp4condition}
\end{equation}
so the fixed line becomes
\begin{equation}
K=2/\pi,
\qquad
y^2-y_4^2=\kappa/(16\pi).
\label{eq:LRp4line}
\end{equation}
Equivalently,
\begin{equation}
y=\sqrt{y_*^2+y_4^2},
\qquad
y_*=\sqrt{\frac{\kappa}{16\pi}}.
\label{eq:LRp4y}
\end{equation}
This line emanates from the pure LR fixed point \((K_*,y_*,0)\), rather than from \((K_*,0,0)\) as in the SR case. 
To linearize around the line, set \(P=y^2\), \(Q=y_4^2\), and \(w=\delta P-\delta Q\). One finds
\[
\frac{d}{d\ell}
\begin{pmatrix}
\delta K\\
w
\end{pmatrix}
=
\begin{pmatrix}
-16\pi^2P & -16\pi\\
-2\pi(P+Q) & 0
\end{pmatrix}
\begin{pmatrix}
\delta K \\ w
\end{pmatrix}.
\]
The eigenvalues satisfy
\( \lambda^2+16\pi^2P\,\lambda-32\pi^2(P+Q)=0\),
and the relevant eigenvalue is
\begin{equation}
\lambda_+ =
-8\pi^2P
+
\sqrt{64\pi^4P^2+32\pi^2(P+Q)}.
\label{eq:LRp4lambda}
\end{equation}
The exponent \(\nu=1/\lambda_+\) depends on both the LR coupling \(\kappa\) and the fourfold field \(h_4\) via the fugacity \(y_4\).

Two limits are useful. For \(y_4\ll y_*\), which for fixed \(\kappa\) corresponds to \(h_4\to0\), we have \(P\simeq\kappa/(16\pi)\), \(Q\simeq0\), and
\[
1/\nu\simeq \sqrt{2\pi\kappa}
\qquad(y_4\ll y_*).
\]
In this limit the fourfold field is a weak perturbation on the pure LR fixed point, and \(\nu\) is controlled by the LR mechanism. 
For \(y_4\gg y_*\), the perturbative expression for \(\nu\) takes the form
\begin{equation}
1/\nu\simeq 8\pi y_4.
\label{eq:LRp4nu2}
\end{equation}
This expression has a clear physical meaning only by taking \(\kappa\to0\) with \(h_4\) small and finite: then \(y_*\to0\), the LR tail disappears, and Eq.~\eqref{eq:LRp4nu2} recovers the SR \(p=4\) result, valid for \(h_4\ll1\). If instead \(\kappa\) is held fixed and \(h_4\) grows to \(O(1)\), the fourfold field locks \(\theta\) to four directions, and the model reduces to a four-state clock model equivalent to two decoupled Ising models, lying outside the present perturbative RG treatment.

The crossover between the two regimes occurs at \( y_4\sim y_*=\sqrt{\kappa/(16\pi)}\).
The low-\(T\) phase for \(h_4\neq0\) remains fourfold long-range ordered.

\subsection{The case \(p>4\): crossover and locking}
\label{sec:pgt4}

For \(p>4\), the \(p\)-fold field is irrelevant at $K_*=2/\pi$, since
\( \lambda_p(K_*)=2-(p^2/8)<0 \).
Consequently,  the transition at \(T_c\) is continuous and is controlled by the pure LR fixed point \((K_*,y_*,0)\), with \( 1/\nu\simeq \sqrt{2\pi\kappa} \).
This is the only thermodynamic transition for \(p>4\), directly from the disordered into the \(p\)-fold LRO phase.

Below \(T_c\), the stiffness exceeds \(K_*\) and the vortex fugacity is exponentially small. The stiffness then grows monotonically,
\( dK/d\ell \simeq\kappa \),
and reaches \( K_p=p^2/(8\pi) \) at a finite scale
\begin{equation}
\ell_c\sim (K_p-K_0)/\kappa.
\label{eq:lc}
\end{equation}
For \(K<K_p\), the \(p\)-fold field remains irrelevant and the system exhibits log-QLRO behavior. For \(K>K_p\), the field becomes relevant and grows; the feedback term \(+\pi p^2y_p^2\) in the stiffness flow accelerates the growth of $K$. The system locks into the \(p\)-fold LRO phase.

Thus, for any finite \(p>4\), the log-QLRO regime is a finite-scale crossover rather than a thermodynamic phase. 
There is no second phase transition at \(K_p\): the locking is a crossover, not a critical phenomenon. 
The crossover scale grows with \(p\). As \(p\to\infty\), \(K_p\to\infty\), the stiffness never reaches \(K_p\) at finite temperature, the locking scale diverges, and the pure XY log-QLRO phase is recovered. This is consistent with the expectation that the \(p\to\infty\) limit recovers the pure XY model. 

The projected \(p>4\) RG flows of Eqs.~\eqref{eq:unifiedK}--\eqref{eq:unifiedy} are shown in Fig.~\ref{fig:p6projectednew}(a) for the SR (\(\kappa=0\)) case and Fig.~\ref{fig:p6projectednew}(b) for the LR (\(\kappa>0\)) case. 
Comparison of Figs.~\ref{fig:purexyflowsnew} and~\ref{fig:p6projectednew} illustrates the opposing stiffness renormalization by vortices and \(p\)-fold anisotropy: whereas vortex softening (\(-4\pi^3K^2y^2\)) tends to bend the flows toward smaller \(K\) in Fig.~\ref{fig:purexyflowsnew}, anisotropic stiffening (\(+\pi p^2y_p^2\)) can bend trajectories toward larger \(K\) in Fig.~\ref{fig:p6projectednew}; the net direction is set by the competition of all terms in Eq.~\eqref{eq:unifiedK}. Moreover, when \(\kappa>0\) [Fig.~\ref{fig:p6projectednew}(b)], the SR fixed segment and \(K_p\) critical endpoint disappear: although ordered-side trajectories appear to hit the \(y_p=0\) axis before reaching \(K_p\) due to exponential decay for \(K<K_p\), \(y_p\) remains nonzero for nonzero initial anisotropy and eventually turns upward past \(K_p\), producing the extended finite-scale log-QLRO crossover.

\subsection{The case \(p<4\): breakdown of the perturbative framework}
\label{sec:plt4}

For \(p<4\), the \(p\)-fold field is relevant at the pure LR fixed point,
\( \lambda_p(K_*)=2-(p^2/8)>0 \).
The small-\(h_p\) perturbative treatment therefore breaks down in the infrared: the field grows under the RG and drives the system away from the pure fixed point. The critical behavior lies beyond the present RG framework, which assumes \(y_p\ll1\) throughout. 
A separate analysis is required.

\section{Conclusions and discussion}
\label{sec:conclusions}

We have analyzed the two-dimensional LR XY model with \(p\)-fold anisotropy at the marginal case \(\sigma=2\), using the physical vortex and \(p\)-fold fugacities and the unified lowest-order RG equations~\eqref{eq:unifiedK}--\eqref{eq:unifiedy}. The main results are summarized in Table~\ref{tab:comparison} 
and the phase diagram is illustrated schematically in Fig.~\ref{fig:phaseproposal}.

\begin{table}[t]
\centering
\caption{Comparison of SR and marginal LR ($\sigma=2$) behavior for pure XY and weak $p$-fold anisotropy, within the lowest-order RG. For LR $p>4$ at nonzero field, log-QLRO is a crossover preceding LRO. The $p<4$ case lies beyond the weak-field perturbative treatment.}
\label{tab:comparison}
\setlength{\tabcolsep}{3pt}
\begin{ruledtabular}
\begin{tabular}{lcc}
 & SR & LR \(\sigma=2\) \\
\hline
\multicolumn{3}{l}{\textbf{Pure XY}} \\
Fixed point & \begin{tabular}{@{}c@{}}$y=0$; \\$2\pi K>4$\end{tabular}
            & \begin{tabular}{@{}c@{}}$(K_*,y_*,0)$\end{tabular} \\
Transition & one BKT & one continuous \\
Intermediate & --- & --- \\
Low-\(T\) & algebraic QLRO & log-QLRO \\
\(\nu\) & BKT & \(1/\sqrt{2\pi\kappa}\) \\
\hline
\multicolumn{3}{l}{\textbf{Case \(p=4\)}} \\
Fixed point & \begin{tabular}{@{}c@{}}$K=2/\pi$; \\$y=\pm y_4$\end{tabular}
            & \begin{tabular}{@{}c@{}}$K=2/\pi$; \\$y^2-y_4^2=\kappa/(16\pi)$\end{tabular} \\
Transition & one continuous & one continuous \\
Intermediate & --- & --- \\
Low-\(T\) & 4-fold LRO & 4-fold LRO \\
\(\nu\) as \(h_4\to0\) & \(\sim1/|h_4|\to\infty\) & \(\to1/\sqrt{2\pi\kappa}\) \\
\hline
\multicolumn{3}{l}{\textbf{Case \(p>4\)}} \\ 
Fixed point & \begin{tabular}{@{}c@{}}$y=0$; \\$4<2\pi K<p^2/4$\end{tabular}
            & \begin{tabular}{@{}c@{}}$(K_*,y_*,0)$\end{tabular} \\
Transition  & two BKT & one continuous \\
Intermediate & algebraic QLRO & log-QLRO crossover \\
Low-\(T\) & \(p\)-fold LRO & \(p\)-fold LRO \\
\(\nu\) & BKT & \(1/\sqrt{2\pi\kappa}\) \\
\end{tabular}
\end{ruledtabular}
\end{table}

Three features stand out. First, for the pure XY model, the SR algebraic QLRO is replaced at \(\sigma=2\) by log-QLRO, and the BKT singularity of \(\nu\) is replaced by the finite power-law value \(1/\sqrt{2\pi\kappa}\); the fixed point moves from the line \(y=0\) to the single point \((K_*,y_*,0)\).
Second, for \(p=4\), the SR fixed line \(y=\pm y_4\) is shifted to \(y^2-y_4^2=\kappa/(16\pi)\), emanating from the pure LR fixed point. Consequently, \(\nu\) depends on both \(\kappa\) and \(h_4\); in the SR case, by contrast, \(\nu\) is a function of \(h_4\) alone.
Third, for \(p>4\), the two SR BKT transitions are replaced by a single continuous transition at \(T_c\), 
and the log-QLRO regime is reduced to a finite-scale crossover before the system locks into the \(p\)-fold LRO phase. There is no second thermodynamic transition.

A subtlety arises in the interpretation of \(\nu\) for \(p>4\). At the level of the RG equations, \(\nu\simeq1/\sqrt{2\pi\kappa}\) appears to depend only on \(\kappa\), but \(\kappa\) is evaluated at \(T_c\), which shifts with \(h_p\); \(\nu\) therefore varies continuously with \(h_p\) even though \(h_p\) does not appear explicitly. This indirect dependence may, however, be numerically small: \(\kappa\propto1/T_c\), and \(T_c\) changes only modestly with \(h_p\) over the accessible range, so detecting the resulting variation of \(\nu\) may require high-precision simulations.

In all three cases studied---the pure XY model, the \(p=4\) case, and the \(p>4\) case---the SR critical behavior is qualitatively changed at \(\sigma=2\), supporting \(\sigma_*=2\) as the boundary between SR and LR universality classes. This is consistent with the \(\epsilon\)-expansion results for LR classical and quantum \(O(N)\) models, and percolation near their upper critical dimensions~\cite{Li2026Ising,Li2026Quantum,Li2026Percolation}, as well as with high-precision simulations of a broad variety of LR systems~\cite{Liu2025,Liu2026Percolation2,Xiao2026LERW,Xiao2025Universality}, all of which support \(\sigma_*=2\) rather than Sak's criterion \(\sigma_*=2-\eta_{\rm SR}\)~\cite{Sak1973}.

In the SR limit, the \(p=4\) fixed line is closely linked to the \(K_4<0\) region of the Ashkin--Teller (AT) model. Along the \(p=4\) line \(\nu\) increases continuously from the Ising value (\(\nu=1\)) at \(h_4\to\infty\) to the BKT value (\(\nu\to\infty\)) at \(h_4=0\); along the AT line the same happens from \(K_4=0\) to \(K_4\to-\infty\). Moreover, the triangular-lattice AT model with \(K_4\to-\infty\) maps onto the honeycomb-lattice \(O(2)\) loop model, both in the BKT universality class. Given this close relation, it is tempting to speculate that the SR--LR boundary \(\sigma_*=2\) holds true for all three families of models.
Direct numerical studies or controlled RG treatments are needed.

There is also a broader methodological perspective. 
The exact critical exponents of many two-dimensional statistical models, including percolation, Ising, Potts, and \(O(N)\) models, can be obtained by mapping them onto Coulomb-gas theories, similar to the XY and \(p\)-fold XY models. Since the \(\sigma=2\) infrared kernel differs from the SR kernel only by the factor \(\ln(1/k)\), the adiabatic procedure may be generalizable to a logarithmic Coulomb gas, which may provide a unified approach to these models at the marginal case \(\sigma=2\).

Several limitations should be noted. The unified equations are the lowest-order terms of an adiabatic expansion in which the LR coupling \(\kappa\) is treated as scale independent. In the critical region, the vortex fugacity is no longer exponentially small and \(\kappa\) itself may flow. The qualitative conclusions are expected to survive these corrections, but the quantitative values of critical parameters should be modified. Therefore, numerical studies, especially large-scale Monte Carlo simulations, remain essential for locating the critical points and for determining the critical exponents quantitatively.

\begin{acknowledgments}
We are grateful to Zhijie Fan for valuable discussions.
We acknowledge support from the National Natural Science Foundation of China (NSFC) under Grants No. 
11975024 and 12275263, as well as the Quantum Science and Technology National Science and Technology Major Project (Grant No. 2021ZD0301900). 
\end{acknowledgments}

\bibliography{refs}
\end{document}